\documentclass[aps,nofootinbib,twocolumn,floatfix,superscriptaddress,secnumarabic]{revtex4-1}
\usepackage{amsfonts,amssymb,epsfig,verbatim,mathbbol,amstext,amsmath,psfrag}
\usepackage{graphicx}
\usepackage{t1enc}
\usepackage{amsmath}
\usepackage{subfigure}
\usepackage{dsfont}
\usepackage{slashed}

\usepackage{colordvi}
\usepackage{xcolor}
\usepackage{color}
\newif\ifSMver
\SMverfalse

\ifSMver
\newcommand{\mysec}[1]{{\em #1} ---}

\else
\newcommand{\mysec}[1]{\section{#1}}

\fi

\usepackage{tikz}
\usetikzlibrary{trees,patterns}
\usetikzlibrary{decorations.pathmorphing}
\usetikzlibrary{decorations.markings}
\usetikzlibrary{shapes.misc}

  \definecolor{jblue}  {RGB}{20,50,100}
  \definecolor{npurple}  {RGB} {153, 51, 204}
  \definecolor{wred}   {RGB}{217,0,56}
  \definecolor{white}   {RGB}{255,255,255}

  \definecolor{korange}   {RGB}{235, 80,  43}
  \definecolor{korange2}   {RGB}{245, 100,  63}
  \definecolor{kyelloworange}   {RGB}{255, 210,  110}
  \definecolor{kyelloworange2}   {RGB}{240, 170,  90}
  \definecolor{kred}   {RGB}{204,  102, 153}
  \definecolor{kpurple}   {RGB}{153,  61, 190}
  \definecolor{kpurplelight}   {RGB}{213,  161, 230}
        \tikzset{
          photon/.style={decorate, decoration={snake}, draw=npurple,very thick},
          boson/.style={decorate, decoration={snake}, draw=npurple,very thick},
          electron/.style={draw=jblue,very thick, postaction={decorate},
                   decoration={markings,mark=at position .55 with {\arrow[draw=jblue]{>}}}
          },
          electron2/.style={draw=jblue,very thick, postaction={decorate},
                   decoration={markings,mark=at position .55 with {\arrow[draw=jblue]{<}}}
          },
          fermion/.style={draw=jblue,very thick, postaction={decorate},
                    decoration={markings,mark=at position .55 with {\arrow[draw=jblue]{}}}
          },
          gluon/.style={decorate, draw=korange,very thick, %kred
            decoration={coil,amplitude=4pt, segment length=6pt}},
          higgs/.style={draw=wred,very thick, postaction={decorate},
                   decoration={markings,mark=at position .55 with {\arrow[draw=wred]{>}}}
          },
          nothing/.style={draw=white,very thick}
        }

\tikzset{cross/.style={cross out, draw=black, minimum size=2*(#1-\pgflinewidth), inner sep=0pt, outer sep=0pt},
cross/.default={1pt}}

\newcommand{\be}{\begin{equation}}
\newcommand{\ee}{\end{equation}}

\newcommand{\dd}{\textmd{d}}

\newcommand{\Z}{\mathcal{Z}}

\newcommand{\expv}[1]{\left \langle #1 \right \rangle}

\newcommand{\tr}{\textmd{tr}\,}

\newcommand{\phimin}{\bar\phi_{\rm min}}

\newcommand{\Bielefeld}{Fakult\"at f\"ur Physik, Universit\"at Bielefeld,
  D-33615 Bielefeld, Germany.}
\newcommand{\Budapest}{E\"otv\"os Lor\'and University, P\'azm\'any P\'eter\ s\'et\'any\ 1/A, H-1117, Budapest
  Hungary.}
\newcommand{\Lendulet}{Institute for Nuclear Research, Bem t\'er 18/c, H-4026 Debrecen, Hungary.}

\usepackage[    bookmarks,
                 bookmarksopen = true,
                 bookmarksnumbered = true,
                 breaklinks = true,
                 linktocpage,
                 colorlinks = true,
                 linkcolor = blue,
                 urlcolor  = blue,
                 citecolor = blue,
                 anchorcolor = green,
                 hyperindex = true,
                 hyperfigures]
        {hyperref}

\begin{document}

\title{
Spontaneous symmetry breaking via inhomogeneities and the differential surface tension
}

\author{G.~Endr\H{o}di}
\affiliation{\Bielefeld}
\author{T.~G.~Kov\'acs}
\affiliation{\Budapest}
\affiliation{\Lendulet}
\author{G.~Mark\'o}
\email[Corresponding author, email: ]{gmarko@physik.uni-bielefeld.de}
\affiliation{\Bielefeld}

\begin{abstract}
We discuss spontaneously broken quantum field theories with a continuous symmetry group via the constraint effective potential. Employing lattice simulations with constrained values of the order parameter, we demonstrate explicitly that the path integral is dominated by inhomogeneous field configurations and that these are unambiguously related to the flatness of the effective potential in the broken phase.
We determine characteristic features of these inhomogeneities, including their topology and the scaling of the associated excess energy with their size. Concerning the latter we introduce the differential surface tension -- the generalization of the concept of a surface tension pertaining to discrete symmetries. Within our approach, spontaneous symmetry breaking is captured merely via the existence of inhomogeneities, i.e.\ without the inclusion of an explicit breaking parameter and a careful double limiting procedure to define the order parameter. While here we consider the three-dimensional $\mathrm{O}(2)$ model, we also elaborate on possible implications of our findings for
the chiral limit of QCD.
\end{abstract}

%\pacs{12.38.Gc, 13.20.Cz, 26.60.-c, 26.50.+x}

\maketitle

\mysec{Introduction}
Spontaneous symmetry breaking is one of the most important general concepts of quantum field theories. It is responsible for prominent features of Quantum Chromodynamics as well as the electroweak sector of the Standard Model, including the characteristics of the spectrum of the theory and, in general, how the mass of the visible matter in our Universe is generated. Via the Peccei-Quinn mechanism, it also provides a possible explanation for the strong CP problem in terms of axions.

While in the standard picture the Higgs and chiral condensates are homogeneous in space, the possibility of inhomogeneous symmetry breaking has been discussed in the literature for both sectors. Impurities in the Higgs condensate~\cite{Hosotani:1982ii} might generally arise via phase transitions in the early universe~\cite{Kibble:1997yn}, through false vacuum decay after inflation~\cite{GarciaBellido:2002aj} or via further mechanisms~\cite{Giudice:2010zb}. An inhomogeneity in the chiral condensate has been discussed in the context of the mismatch between the contribution to the vacuum energy and experimental constraints on the cosmological constant~\cite{Brodsky:2008xu,Brodsky:2010xf,Brodsky:2012ku}. In the QCD sector, further types of inhomogeneities might emerge,
e.g.\ the so-called chiral spiral,
see the recent review~\cite{Buballa:2014tba}. These are expected at high chemical potentials, as predicted
by low-dimensional soluble models~\cite{Schon:2000he,Basar:2009fg} and numerical simulations~\cite{Lenz:2020bxk,Wipf_to_appear} or in a possible quarkyonic phase~\cite{Kojo:2009ha}. Inhomogeneous axion fields might lead to the formation of so-called axion miniclusters, see e.g.~\cite{Marsh:2015xka}.

Here we do not intend to investigate such inhomogeneities induced by
external parameters like a chemical potential. Instead, we will discuss the
inhomogeneity of order parameters in the general context of
spontaneous symmetry breaking in quantum field theories with a continuous symmetry group.
The standard approach to construct the order parameter amounts to including an explicit symmetry breaking parameter $h$ in an arbitrary direction in internal space and performing the limit $h\to 0$ in the infinite volume. The so defined order parameter is homogeneous and approaches the value $\phimin$ from above in this double limit. In contrast, our present approach -- based on the constraint effective potential~\cite{ORaifeartaigh:1986axd} -- operates directly at $h=0$ and enables exploring arbitrary values of the average order parameter $\bar\phi$ with $|\bar\phi|<\phimin$.

In this region the local order parameter is anticipated to be inhomogeneous, in line with the flatness of the effective potential.
This can be understood simply for {\it discrete} symmetries. Since the local order parameter prefers amplitudes around $\phimin$, averages below this value are only possible if
the field occupies one discrete minimum in one fraction of the volume and another one in the rest~\cite{PhysRevD.36.2474}. Such bubbles carry an action that scales with their
surface, becoming negligible compared to volume averages in the thermodynamic limit,
ensuring the flatness of the effective potential.
The formation of bubbles may be thought of as a natural consequence of the first-order nature of the phase transition that takes place in the broken phase as $h$ crosses zero and the order parameter flips sign.

Similar inhomogeneities are expected to appear
for $|\bar\phi|<\phimin$ for {\it continuous} symmetries as well.
In this case the different orientations of the field in internal space will be connected continuously by spin-wave-like configurations. In this paper we discuss how such inhomogeneities emerge in the
constrained path integral
and investigate their properties,
uncovering finer details of the hidden structure of the flat potential.
Furthermore, employing the excess energy associated to inhomogeneous configurations,
we extend the definition of a surface tension from the bubble formation in the discrete case to systems with continuous symmetry,
and provide first results for it.
Our simulations are carried out in the three-dimensional
$\mathrm{O}(2)$ model, but similar results are expected in general
for continuous, spontaneously broken systems.

\mysec{Spontaneous symmetry breaking}
We consider the three-dimensional $\mathrm{O}(2)$ model
involving the scalar field $\phi_a(x)$ ($a=0,1$),
described by the Lagrangian density
\begin{equation*}
\mathcal{L}(x)= \frac{m^2}{2}\sum_a\phi_a^2(x) +\frac{g}{24}\Big[\sum_a\phi_a^2(x)\Big]^2\!+\,\frac{1}{2}\sum_{\mu,a}[\partial_\mu \phi_a(x)]^2 \,,
\end{equation*}
with the coupling $g$ and the squared mass $m^2$, which we
take to be negative. For homogeneous fields, $\phi_a(x)=\bar\phi_a$, the Lagrangian
equals the
spontaneously breaking classical potential $\Omega_{\rm cl}(\bar\phi)$ taking its minimum along
the valley $|\bar\phi|=\phimin^{\rm cl}$ and being concave on the
disk inside that.
The {\it canonical} partition function reads
\be
Z_h=\int [\dd\phi_a(x)] \exp\left\{-\int \!\dd^3 x \left[ \mathcal{L}(x)-h\,\phi_0(x)\right]\right\}\,,
\label{eq:Z}
\ee
where we also included an explicit symmetry breaking term, proportional to
the magnetic field $h$,
chosen to point in the $0$ direction in internal space without
loss of generality.

In the direction specified by the explicit symmetry breaking parameter we have
\be
\expv{\phi_0}_h = \frac{1}{V}\frac{\partial \log Z_h}{\partial h}\,,
\label{eq:phi0}
\ee
where the expectation value is understood according to
the partition function $Z_h$ from Eq.~(\ref{eq:Z}) and $V$ denotes
the volume of the system.
In the $h\to0$ limit $\expv{\phi_0}_h$ vanishes in finite volumes.
The standard approach to define the order parameter at
the symmetric point is via the double limit procedure
\be
\phimin = \lim_{h\to0} \lim_{V\to\infty} \expv{\phi_0}_h\,,
\label{eq:phibarmin}
\ee
often found in textbooks. In this approach, values of the average
order parameter inside the disk
$|\bar\phi| <\phimin$
are inaccessible. We will refer to this as the coexistence disk.

Next we consider an alternative approach based on the {\it constrained} path integral,
\be
\Z_{\bar\phi}\!=\! \int [\dd\phi_a(x)]\, e^{-\int\!\dd^3x\, \mathcal{L}(x)}\,\delta^{(2)}\!\left(\frac{1}{V}\!\int\!\dd^3 x\, \phi_b(x)-\bar\phi \,\delta_{b0}\!\right),
\ee
and the constraint potential $\Omega(\bar\phi)=-\log\Z_{\bar\phi}/V$ associated to it. Again without loss of generality, we can choose the constrained field to lie in the $0$ direction.
Expectation values according to $\Z_{\bar\phi}$ will be denoted
by $\expv{\,}_{\bar\phi}$. Most importantly, the derivative with respect to $\bar\phi$ gives the magnetic field as an observable,
via the explicit formula~\cite{Fodor:2007fn},
\be
\label{eq:hdef}
\expv{h}_{\bar\phi}
\equiv \frac{\partial \Omega(\bar\phi)}{\partial \bar\phi}
= m^2 \bar\phi + \frac{g}{6V} \expv{\int\!\dd^3x \sum_{a}\phi_a^2(x)\phi_0(x)}_{\!\bar\phi}\,.
\ee

A related quantity is the Legendre transform of the canonical free energy
\be
\Gamma(\bar\phi) \equiv \underset{h}{\rm sup} \left(h\cdot \bar\phi - \frac{1}{V} \log Z_{h}\right)\,.
\label{eq:Gammadef}
\ee
It is well known that in the thermodynamic limit the two potentials are equal: $\Omega(\bar\phi)=\Gamma(\bar\phi)$~\cite{ORaifeartaigh:1986axd,Kuti:1987bs}. Moreover, $\Gamma(\bar\phi)$ is convex and, in particular,
flat on the coexistence disk in the infinite volume limit. (Even though here $\Gamma$ and $\Omega$ are functions of a scalar variable,
they can be extended to the $\mathrm{O}(2)$ plane via their
invariance under rotations.)
Carrying out the supremum in Eq.~\eqref{eq:Gammadef} also defines $h(\bar\phi)$, whose inverse equals the function $\expv{\phi_0}_h$.
From the equality of the two potentials it
also follows that
$\expv{h}_{\bar\phi}=h(\bar\phi)$ holds
in the thermodynamic limit.

Simulations are, in turn, performed in a finite volume. While $\Gamma$ retains its convexity by construction, $\Omega$ can be concave for $V<\infty$. How the convexity is recovered as the volume increases was already discussed in~\cite{ORaifeartaigh:1986axd}. A more specific question is what type of configurations dominate the path integral as the volume increases. As we already argued above, we expect to see spin-wave-like configurations for $|\bar\phi|<\phimin$.

\mysec{Simulation results\label{sec:results}}
We perform constrained simulations of the three-dimensional $\mathrm{O}(2)$ model with parameters $m^2=-15.143$ and $g=102.857$, corresponding to the broken symmetry phase. The simulated values of $\bar\phi$ range over the full coexistence disk and slightly outside of it on three different volumes $40^3,\,60^3$ and $80^3$. We employed two algorithms: first, a Metropolis-like approach, where randomly chosen pairs of spins are updated so as to keep $\bar\phi$ fixed and second, a constrained hybrid Monte Carlo algorithm based on Ref.~\cite{Fodor:2007fn}. The two algorithms were cross-checked and gave the same result.

We first examine the typical configurations that dominate $\Z_{\bar\phi}$. Since the field is strongly constrained in the radial direction by the classical potential, its magnitude is expected to stay close to $\phimin$ and a constrained expectation smaller than that can only be achieved by inhomogeneous, spin-wave type configurations. Indeed,
starting from homogeneous initial states $\phi_a(x)=\bar\phi\, \delta_{a0}$, our constrained simulations are quickly driven towards inhomogeneous configurations.
The direction of the emerging, coherently rotating spin waves is
selected spontaneously so that the associated momentum is
minimal. For our cubic lattices with periodic boundary conditions, this prefers
one of the coordinate axes which, without loss of generality~\cite{z3analogy}, can be taken to be $x_1$.
In the two-dimensional slices perpendicular to $x_1$, the direction of the field has only random fluctuations, thus the nontrivial change of the field angle is restricted to the $x_1$ direction. This implies the spontaneous breakdown of the translational and rotational symmetry of the system -- we stress, however, that expectation values of the field respect both of these symmetries as the waves can be shifted and rotated without changing the action.
The spontaneous breaking of the spacetime-symmetries could be explicitly seen by simulating systems that are slightly elongated in the $x_1$ direction or by including magnetic fields localized to a single $x_1$-slice. Defining the theory via the limit where such small explicit breakings are gradually removed, in the thermodynamic limit, selects one minimum, just as the $h\to0$ limit of the canonical theory selects one $\bar\phi$ orientation.

\begin{figure}[t]
\includegraphics[width=0.45\textwidth]{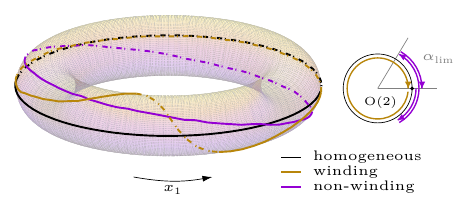}
\caption{\label{fig:torus} Visualization of the typical configurations dominating the constrained path integral for $\bar\phi<\phimin$ in comparison to a homogeneous one. The curves on the torus represent the vector $\Sigma$ from Eq.~\eqref{eq:slicedef} in $\mathrm{O}(2)$ space as a function of the $x_1$ coordinate. All lines are solid (dashed) when they are in front of (behind) the torus. Notice
that the winding configuration is topologically different from the homogeneous one, while the non-winding one is topologically equivalent to it.}
\end{figure}

We characterize the dominant configuration types based on their behaviour in the inhomogeneous direction. Defining slices of the field in the previously specified $x_1$ direction as
\be
\label{eq:slicedef}
\Sigma_a(x_1)\!=\!L^{-2}\int \!\dd^3 y \,\phi_a(y) \,\delta(y_1-x_1)\,,
\ee
configurations can be characterized through the smooth precession of the $\mathrm{O}(2)$ vector $\Sigma$ as a function of $x_1$, while its length remains very close to $\bar\phi_{\rm min}$. In particular, we assign an integer winding number $w$ to each configuration based on how many times the $\mathrm{O}(2)$ group is mapped to the circle
corresponding to the periodic $x_1$ coordinate. We find that while higher $w$-s are also possible, the dominant contributions to the path integral have either $w=0$ or $w=1$. Since practically only these two are present we will refer to $w=1(0)$ as (non-)winding configurations. To facilitate the understanding of this topological classification,
example configurations taken from simulations at $\bar\phi\approx0.5$ in $V=80^3$ are shown and compared to a homogeneous one in Fig.~\ref{fig:torus}.

For $w=1$, the angle $\alpha(x_1)={\rm atan\,}(\Sigma_1(x_1)/\Sigma_0(x_1))$ rotates around the complete internal space (with non-constant velocity), while for $w=0$
it oscillates between the limiting angles $\pm\alpha_{\rm lim}$, see Fig.~\ref{fig:torus}. The constraint that the average field be equal to $\bar\phi$ is therefore achieved in different ways in the two cases. The exact functional form of $\alpha(x_1)$ depends strongly on $\bar\phi$ for any $w$. We use an ansatz for $\alpha(x_1)$ motivated by one-dimensional solutions of the classical equations of motion, which can be written as
\begin{subequations}
\label{eq:ansatz}
\begin{align}
\phi^{(w)}(x_1) &= \phimin\cdot (\cos\alpha_w(x_1),\sin\alpha_w(x_1) )^\top\,,\\
\alpha_w(x_1) &= \frac{2\pi w x_1}{L} + \alpha_{\rm lim} \sin\left(\frac{2\pi x_1}{L}\right)\,.
\end{align}
\end{subequations}

The two sectors are found to be dominant in different regions of the parameter space: $w=0$ is the relevant configuration type near the edge of the coexistence disk,
while $w=1$ becomes dominant near $\bar\phi=0$. At some intermediate value $\bar\phi_c$, a very sharp (already for moderate volumes) transition takes place between the sectors. Due to their topological difference, the sub-dominant sector is metastable and, accordingly, jumps between the two sectors take long in Markov time, even when the difference between the corresponding actions is large. This renders the discussion of the transition between the sectors at $\bar\phi_c$ difficult and prompted us to measure observables in fixed topological sectors. This algorithm therefore works only outside the immediate vicinity of $\bar\phi_c$.

\begin{figure}
\includegraphics[width=0.45\textwidth]{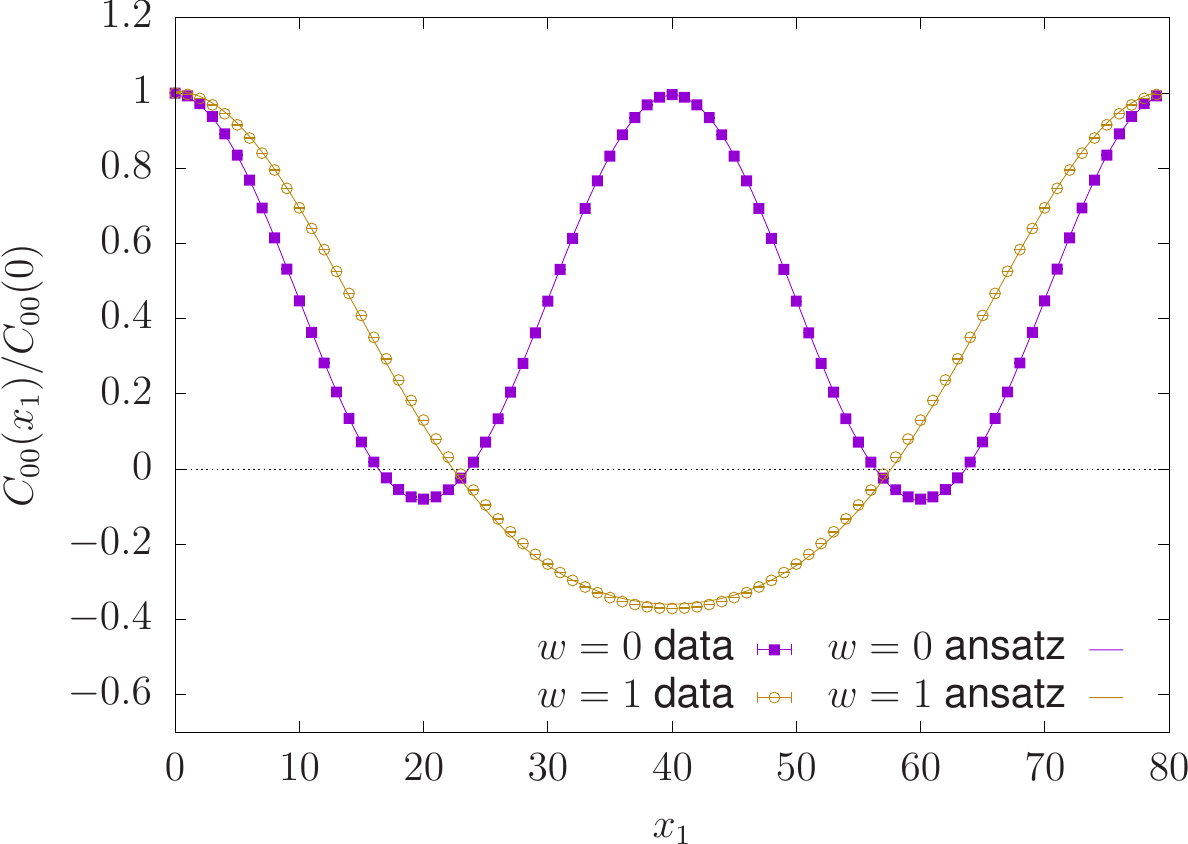}
\caption{\label{fig:corrs} Slice correlators as measured in the winding and non-winding sectors for $V=80^3$, at $\bar\phi\approx0.25$. The respective lines following closely the two datasets are the correlators calculated from the classical ansatz, where the parameter $\alpha_{\rm lim}$ was optimized in a least squares fashion.}
\end{figure}

Owing to translational invariance on the level of expectation values, $\expv{\phi_a(x)}=\expv{\phi_a}$, inhomogeneities are invisible in one-point functions but
may be observed in two-point functions of the field. In particular, we consider the slice correlator,
\be
C_{ab}(x_1)=\expv{\Sigma_a(x_1)\Sigma_b(0)}_{\bar\phi}\,.
\ee
We show examples of $C_{00}$ measured on both topologies in Fig.~\ref{fig:corrs}. A comparison to the correlator calculated from the ansatz of Eq.~\eqref{eq:ansatz} reveals that the latter approximates the simulation results remarkably well. We note that only the non-winding configurations can be continuously connected to the homogeneous ones, which can be seen as the limit $\alpha_{\rm lim} \to 0$~.

\begin{figure}[t]
\includegraphics[width=0.45\textwidth]{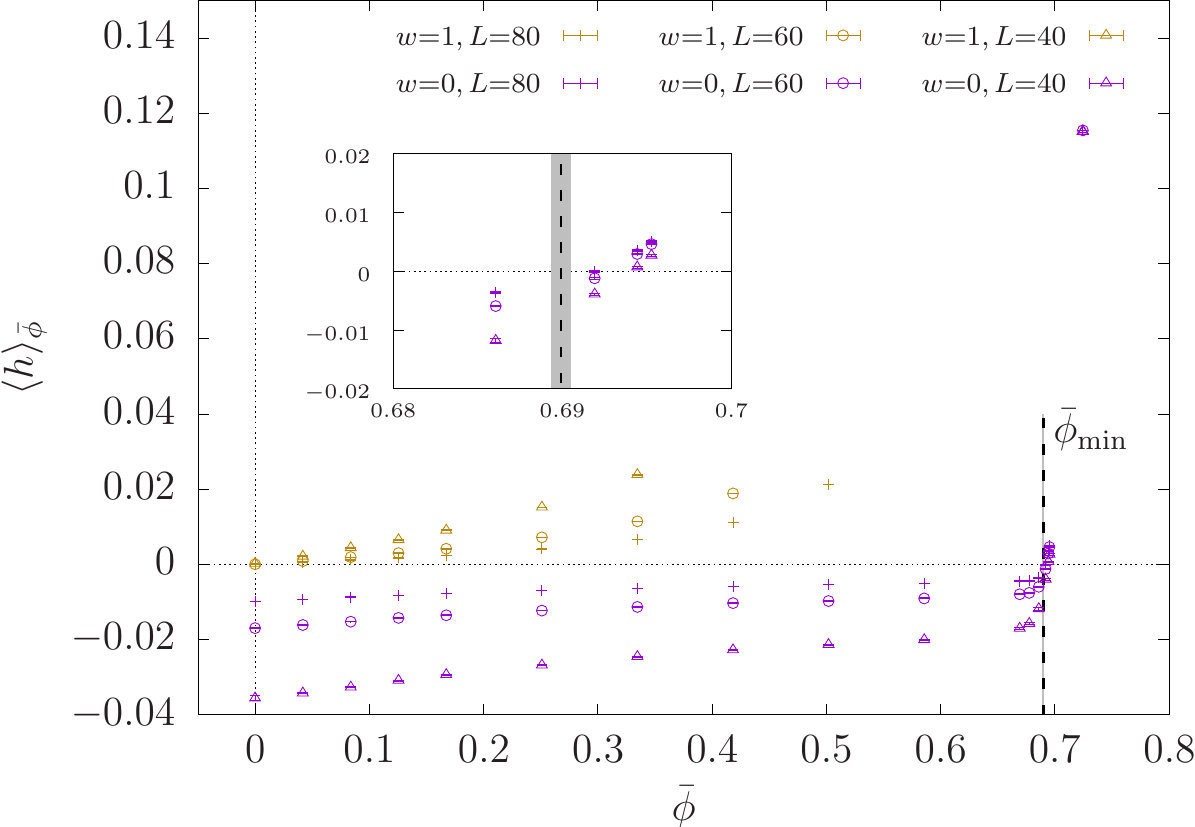}
\caption{\label{fig:h_of_phi} The constrained simulation results for $\langle h\rangle_{\bar\phi}$ as a function of $\bar\phi$. The red (blue) points correspond to measurements on winding (non-winding) configurations. Also shown is the thermodynamic limit extrapolation of $\phimin$ obtained from the intersect of the results with $\langle h\rangle_{\bar\phi}=0$. The inset zooms
into the region around $\phimin$.}
\end{figure}

Next, we investigate how the presence of inhomogeneities is related to the flatness of the effective potential $\Omega(\bar\phi)$. Since the potential itself cannot be expressed as an observable, we determine it via the the so-called integral method. First, we measure the magnetic field $\expv{h}_{\bar\phi}$ based on Eq.~\eqref{eq:hdef}.
This is done separately for the two relevant winding number sectors and the respective observables are marked by a $w$ subscript.
The results are shown in Fig.~\ref{fig:h_of_phi}, revealing how the magnetic field approaches zero in the thermodynamic limit from opposite directions for $w=0$ and $w=1$. The edge of the coexistence disk is determined by the point where $\expv{h}_{\bar\phi,0}$ turns positive and in the thermodynamic limit we obtain $\phimin=0.6899(6)$.

Having measured $\expv{h}_{\bar\phi,w}$, next we integrate it back in $\bar\phi$ to reconstruct the constraint potential via Eq.~\eqref{eq:hdef},
\be
\label{eq:omegaint}
\Omega_{w}(\bar\phi) = \int_0^{\bar\phi} d\bar\phi' \,\expv{h}_{\bar\phi',w} + c_{w}\,,
\ee
where $c_w$ is an integration constant to be set later.
As we explained above, not too close
to $\bar\phi_c$, the potential equals either $\Omega_0$ or $\Omega_1$, whichever is smaller. Since the configurations of the $w=0$ sector connect continuously to the homogeneous configurations at $\phimin$, we can simply set $\Omega_{0}(\phimin)=0$, which fixes the value of $c_0$.

\begin{figure}[t]
\includegraphics[width=0.45\textwidth]{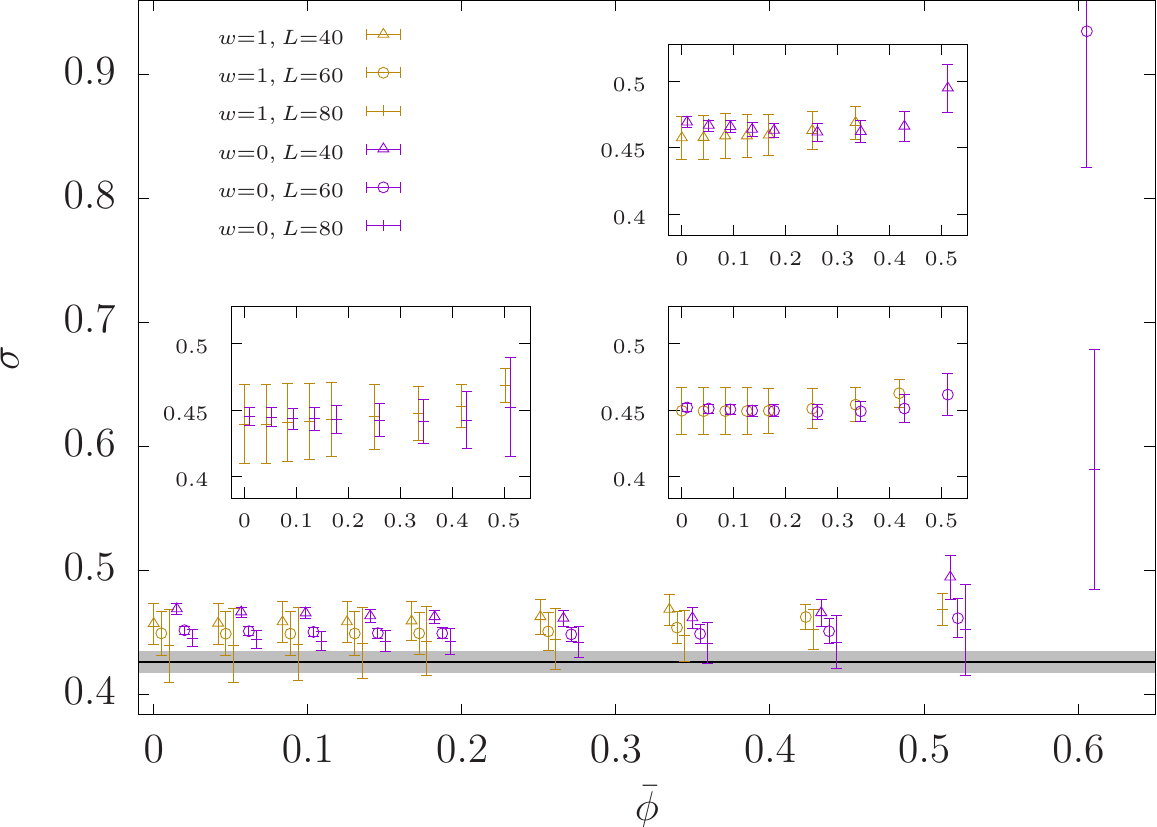}
\caption{\label{fig:surftens} The differential surface tension as defined in Eq.~\eqref{eq:surfacetension_formula} for three different volumes in both relevant topological sectors, together with the
infinite volume limit (gray band). The insets show each volume separately to reveal more clearly the independence of $\sigma$ of the topological sector. Some of the points are shifted horizontally for better visibility.}
\end{figure}

For the winding case this approach is not viable and we opted for an
alternative method. To explain it, we first introduce a generalization of the surface tension, well known in the context of bubble formation for discrete symmetries. There it is defined via the excess free energy of a two-phase configuration containing a bubble wall, compared to a homogeneous, one-phase configuration, per unit wall surface. It characterizes the time scales and bubble nucleation rates at a first order phase transition, see e.g.~\cite{Fraga:2018cvr}.
We can generalize this concept for our continuous symmetry via the kinetic energy of the sliced fields (remember that the spin waves are assumed to point in the $x_1$ direction),
\be
E_\Sigma(\bar\phi) = \frac{1}{2L}\sum_{a,x_1}\expv{[\partial_{x_1} \Sigma_a(x_1)]^2}_{\bar\phi}\,.
\ee
and the excess potential density due to the inhomogeneity at $|\bar\phi|<\phimin$, parameterized as
\be
\Omega(\bar\phi)= \sigma \cdot \frac{E_\Sigma(\bar\phi) -E_\Sigma(\bar\phi_{\rm min})}{\phimin^2}\,,
\label{eq:surfacetension_formula}
\ee
where we used that we set $\Omega(\bar\phi_{\rm min})=0$, hence no subtraction is necessary on the left hand side. In the case of a discrete symmetry with a domain wall of characteristic width $\Delta$ separating phases with $+\phimin$ and $-\phimin$, the same formula gives $V\Omega = \sigma \cdot L^2$, where $\sigma\propto 1/\Delta$ and the proportionality factor depends on the precise profile of the wall. Thus, Eq.~(\ref{eq:surfacetension_formula}) properly generalizes the notion of a surface tension to continuous symmetries and we will
refer to it as {\it differential surface tension}. Since $\sigma$ is defined via average quantities, we expect it to be insensitive to the precise form of the preferred spin-waves and, thus, independent of $\bar\phi$.
Our results for the differential surface tension are shown in Fig.~\ref{fig:surftens},
which demonstrates the independence of $\sigma$ of $w$ and of $\bar\phi$, apart from the region close to $\bar\phi_{\rm min}$, where the ratio it involves
becomes of the $0/0$ type and error bars blow up. Incorporating the slight downward trend towards the infinite volume limit, we obtain $\sigma=0.427(8)$.

\begin{figure}
\includegraphics[width=0.45\textwidth]{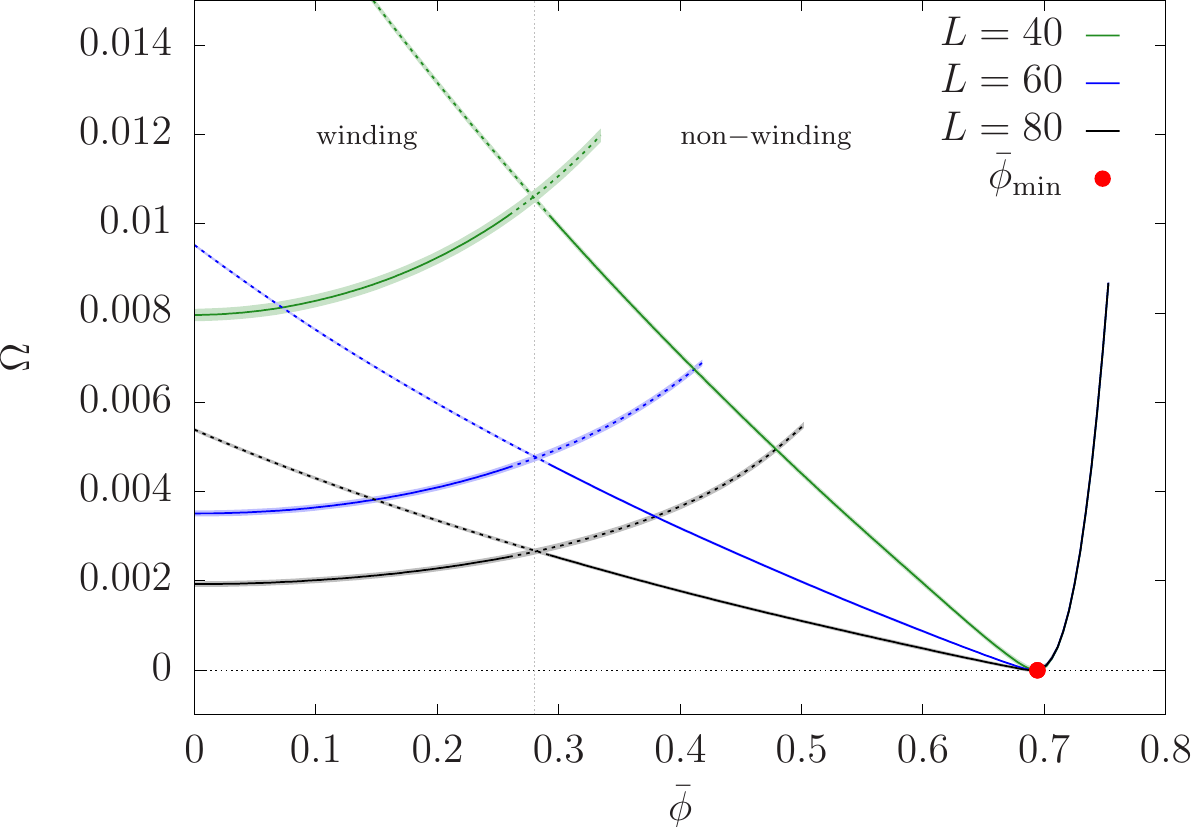}
\caption{\label{fig:pots} The constraint potential for different volumes. The larger the volume the more it approaches the flat effective potential of the thermodynamic limit.}
\end{figure}

Owing to the constancy of $\sigma$, the parameterization Eq.~\eqref{eq:surfacetension_formula} enables us to match the effective potentials measured in the two topological sectors. Since $\Omega_0=\Omega_1$ at $\bar\phi_c$ and $\sigma$ is
the same constant for both, the sliced kinetic energies must also be equal here. Identifying the point where this happens fixes the overall value of $\Omega_1(\bar\phi_c)$ compared to $\Omega_0(\bar\phi_c)$ and hence that of $c_1$.
The so obtained potentials are shown for three different volumes in Fig.~\ref{fig:pots}. The thermodynamic limit of $\Omega$ (assuming $1/L^2$ scaling) is found to be consistent with zero for all values of $\bar\phi$, demonstrating the flatness of the effective potential.
However, for all volumes the cusp associated to the abrupt change of dominant configurations remains. According to our estimate this occurs at $\bar\phi_c=0.2818(2)$.

\mysec{Conclusions}
In this paper we discussed the spontaneously broken phase of the three-dimensional $\mathrm{O}(2)$ model using the constraint
effective potential $\Omega(\bar\phi)$. For large values of $\bar\phi$ this approach reproduces the findings of standard simulations at fixed magnetic field $h$. In addition, inside the coexistence disk ($\bar\phi<\phimin$), it reveals hidden structures of the effective potential and gives insights about the mechanism responsible for flattening towards the thermodynamic limit and the physical realization of infinite ground states in thermodynamics~\cite{ruelle1999statistical}.
In particular, we find that the relevant configurations resemble spin waves labeled by a winding number $w$ according to their $\mathrm{O}(2)$
topology.
Even though the corresponding wavelengths scale with the linear size of the system, a sharp transition takes place between the $w=0$ and $w=1$
sectors at an intermediate critical value $\bar\phi_c$ on each volume.
An algorithm capable of efficiently between the two sectors is yet to be developed.
Incidentally, we point out that our setup provides a counter-example for the
conjecture~\cite{Splittorff:2000mm} that the breaking of translational
invariance is only possible for path integrals with a sign problem.

In finite volumes the constraint potential is concave due to the
excess energy carried by the spin waves. This can be parameterized by a differential surface tension $\sigma$, generalizing the concept of a usual surface tension relevant for discrete symmetries. We provide a first determination of $\sigma$ and show that it is a local property of the quantum field related to its response to torsion and does not depend on the global structure of the waves. These findings might be relevant for phenomenological studies of the impact of inhomogeneous structures for phase transitions, similar to the discussion of
bubble formation for discrete symmetries.
We note that similar inhomogeneities can also be discussed employing boundary conditions
instead of constraints, see e.g.~\cite{Delfino:2018bff,Panero:2020eow}.

It is also worth mentioning that our constrained simulations enable
an efficient determination of $\phimin$ involving an interpolation
in $\bar\phi$ followed by an extrapolation to the thermodynamic limit -- as opposed to the double extrapolation procedure of the standard approach.
Our method can also be applied to QCD in the chiral symmetry
broken phase in $3+1$ dimensions, where the roles of the order parameter and the magnetic field are played by the chiral condensate
and the light quark mass, respectively.
In this case the path integral
will be dominated by configurations with inhomogeneous chiral condensates
inside the coexistence disk. Their specific structure, topological properties and associated differential surface tensions can be
investigated using the methods we introduced in the present work.

\iffalse
We demonstrate this last point explicity by
evaluating the connected part $\chi^{\rm c}$ of the
chiral susceptibility. After the fermionic path integral,
this reads
\be
\chi^{\rm c} = -\frac{T}{V}\expv{\tr (\slashed{D}+m)^{-2}}
= \int_0^\infty \!\!\dd \lambda \, \frac{2(\lambda^2-m^2)}{(m^2+\lambda^2)^2}\, \expv{\rho(\lambda)}\,,
\ee
where we employed the eigensystem $\slashed{D} \psi_\lambda=i\lambda \psi_\lambda$ and $\rho$ denotes the density of eigenvalues.
The $\lambda$-independent part of $\expv{\rho}$ does not
contribute to the integral as can be checked explicitly. The slope of
$\expv{\rho(\lambda)}$ at the origin, however, gives a
contribution that is logarithmically divergent in the infrared.\footnote{A similar argument was also employed to show the infrared
divergence of
the two-point function of
flavor-nonsinglet scalar currents~\cite{Smilga:1993in}.}
\fi

\noindent
{\em Acknowledgments}.
This research was funded by the DFG (Emmy Noether Programme EN 1064/2-1 and the
the Collaborative Research Center CRC-TR 211 ``Strong-interaction
matter under extreme conditions'' -- project number
315477589 - TRR 211). The authors thank Bastian Brandt, D\'aniel N\'ogr\'adi, Attila P\'asztor and Andreas Wipf for insightful comments.

\bibliography{flattening}
\bibliographystyle{utphys}

\end{document}